\documentclass{article}

 \PassOptionsToPackage{numbers, compress}{natbib}
\usepackage[final]{nips_2016}

\usepackage{xr-hyper}

\usepackage[utf8]{inputenc} 
\usepackage[T1]{fontenc}    
\usepackage{hyperref}       
\usepackage{url}            
\usepackage{booktabs}       
\usepackage{amsfonts}       
\usepackage{nicefrac}       
\usepackage{microtype}      

\usepackage[parfill]{parskip}

\usepackage{alex}
\usepackage{url, bm, color}
\usepackage{subfigure}

\usepackage{amsmath,amssymb,amsthm,bm}
\usepackage{algorithm}
\usepackage[noend]{algpseudocode}

\usepackage{multibib}
\newcites{app}{References}

\usepackage[english]{babel}
\usepackage{mdwlist}

\usepackage{algorithm}

\usepackage{comment}

\makeatletter
\def\BState{\State\hskip-\ALG@thistlm}
\makeatother

\newtheorem{theorem}{Theorem}

\newtheorem{remark}{Remark}

\usepackage{xspace}

\usepackage{mathtools}

\usepackage{enumitem}
\usepackage{appendix}

\usepackage{multirow}

\def\P{\mathbb{P}}

\def\R{\mathbb{R}}

\def\cF{\mathcal{F}}

\def\cL{\mathcal{L}}

\def\cT{\mathcal{T}}

\def\TV{\mathrm{TV}}

\title{Joint Hacking and Latent Hazard Rate Estimation}

\author{
  Ziqi Liu \\
  Department of Computer Science and Technology \\
  Xi'an Jiaotong University \\
  \texttt{ziqilau@gmail.com} \\
  \And
  Alexander J.~Smola \\
  Machine Learning Department \\
  Carnegie Mellon University \\
  \texttt{alex@smola.org} \\
  \And
  Kyle Soska \\
  Department of Electrical and Computer Engineering \\
  Carnegie Mellon University \\
  \texttt{ksoska@cmu.edu} \\
  \And
  Yu-Xiang Wang \\
  Machine Learning Department \\
  Carnegie Mellon University \\
  \texttt{yuxiangw@cs.cmu.edu} \\
  \And
  Qinghua Zheng \\
  Department of Computer Science and Technology \\
  Xi'an Jiaotong University \\
  \texttt{qhzheng@mail.xjtu.edu.cn} \\
}

\begin{document}

\maketitle

\begin{abstract}
  In this paper we describe an algorithm for predicting the websites at risk in a
  long range hacking activity, while jointly inferring the provenance
 and evolution of vulnerabilities on websites over continuous time. 
 Specifically, we use hazard regression with a time-varying additive hazard function
 parameterized in a generalized linear form. The activation coefficients on
 each feature are continuous-time functions constrained with total variation
 penalty inspired by hacking campaigns. We show that the optimal solution is
 a $0$th order spline with a finite number of adaptively chosen knots, and
 can be solved efficiently. Experiments on real data show that our method
 significantly outperforms classic methods while providing meaningful interpretability.
\end{abstract}

\section{Introduction}
Websites get hacked, whenever they are subject to a vulnerability that
is known to the attacker, whenever they can be discovered efficiently,
and, whenever the attacker has efficient means of hacking at his
disposal. This combination of \emph{knowledge}, \emph{opportunity},
and \emph{tools} is quite crucial in shaping the way a group of sites
receives unwanted attention by hackers. Unfortunately, as an
observer we are not privy to either one of these
three properties.

Exploits are first discovered by highly skilled hackers who will
use them for their own purposes for an extended
period of time, as long as there is an ample supply of hackable sites
that can be discovered efficiently. Once the \emph{opportunity} for
such hacks diminishes due to exhausted supply, the appropriate
vulnerabilities are often published. The available tools increase
and they are added to the repertoire of
popular rootkits, at the ready disposal of 'script kiddies' who will
attempt to attack the remaining sites. The increased availability of
\emph{tools} often offsets the reduced \emph{opportunity} to yield a
secondary wave of infections. 

In a nutshell, the above leads to the following statistical
assumptions on how vulnerability of sites and the infectious behavior
occurs. Firstly, sites are only practically vulnerable once a
vulnerability is discovered. Second, as time passes, the propensity
of an attack might increase or
not, but changes in attack behavior are discrete
rather than gradual.

We propose a novel hazard regression model that
provides a clear description of the probability a site getting
hacked conditioned on its time-varying features,
therefore allowing prediction tasks
such as finding websites at risk, or inferential tasks such as
attributing attacks to certain features as well as identifying
change points of the activations of certain features to be
conducted with statistical rigor.

\paragraph{Related work.} The primary strategy for identifying web-based malware has been to detect an active infection based on features such as small iFrames~\cite{Provos:Security08}. This approach has been pursued by both academia (e.g.~\cite{Borgolte:CCS13,Invernizzi:S&P12}) and industry (e.g.~\cite{safebrowsing,siteadvisor,NortonSafeWeb}). \citet{Soska2014} propose a data driven (linear classification) approach to identify software packages that were being targeted by attackers to predict the
security outcome of websites.

\section{Hazard Regression}

Hazard regression aims to estimate the chances of survival of a particular
event with covariates $x$, as a function of time, such as to better
understand the effects of $x$. Instead of directly modeling the
cumulative survival distribution, people are
interested in the instantaneous rate of dying of any $x$ at any given
time $t$, i.e. hazard rate $\lambda(x,t)$. The density of dying
at time $t$ is given by 
$\label{eq:hazarddensity}
p(t|x) = \lambda(x,t)\underbrace{p\rbr{\text{survival until $t$}|x}}_{F(t|x)}.
$
This leads to a differential equation for the survival probability
with solution
$
\label{eq:hazardintegral}
F(t|x) = \exp\rbr{-\int_0^t \lambda(x,\tau) d\tau}.
$

In our case, death amounts to a site being infected and $\lambda(x,t)$
is the rate at which such an infection occurs. An extremely useful
fact of hazard regression is that it is additive. That is, if there
are two causes with rates $\lambda$ and $\gamma$ respectively,
it allows us to add the rates and this leads to 
$
F(t|x) = \exp\rbr{-\int_0^t \lambda(x,\tau) + \gamma(x,\tau) d\tau}
$
and
$p(t|x) = \sbr{\lambda(x,t) + \gamma(x,t)} F(t,x)$.
The reason why this is desirable in our case follows from the fact
that we may now model $\lambda$ as the sum of attacks in a generalized
linear form. 

Most hazard regression approaches are based on the Cox's
propotional hazard model
$\lambda(t|x)=\lambda_0(t) \exp(w^\top x)$~\cite{Cox1972},
including parametric models, and nonparametric
models with baseline hazard rate $\lambda_0(t)$ unspecified~\cite{Cox1975}.
The proportional assumption may not hold because of the time-varying
effect of covariates~\cite{buchholzapproaches}. As a result,
time-dependent effect models that allow
$w(t)$ as functions over time for each feature are proposed.
Typically people developed time functions based on fractional polynomials
~\cite{sauerbrei2007new}, or spline functions~\cite{Kooperberg1994}.
Due to the huge parameter space, techniques like reduced rank
methods~\cite{perperoglou2013reduced} and structured penalized methods
~\cite{Tibshirani1997,verweij1995time,perperoglou2014cox} are studied.
However those works either search for global smoothing functions
or need to pre-specify knots. Typically they work on tens of features.
Our work inspired by hacking campaigns
aims to identify discrete attack behaviors. We show the optimal solutioin
is a 0th order spline with knots adaptively chosen over continuous time.

\section{Attributing Hacks}

\subsection{Additive hazard function and variational maximum likelihood}\label{sec:stat}
The blacklist may not always
immediately discover whether a site has been taken over. The
probability that this happens in some time interval $[t_i, T_i]$ is
given by $F(t_i, x_i) - F(T_i, x_i)$, named ``interval censoring''.
On another hand, the absence of
evidence of an infection does not mean the evidence of absence.
In other words, all we know is that the site survived until
time $T$, named ``right censoring''. The
probability is thus given by $F(T,x_i)$.
Time $T$ denotes the end of the observation. Given intervals
$[t_i, T_i]$ of likely infection for site $i \in \{1,...,n\}$,
at time $T$ we have the
following likelihood for the observed data:
\[\prod_{i \in \mathrm{hacked}} \sbr{F(t_i,
	x_i) - F(T_i, x_i)} \prod_{i \not\in \mathrm{hacked}} F(T,x_i).
\]
It remains to specify the hazard function $\lambda(x,t)$.
We do not wish to make strong parametric assumptions, but since $x \in \R^d$ 
is high-dimensional, estimating $\lambda(x,t)$ completely 
non-parametrically is intractable. We thus make an additive
assumption and expand the hazard function into an inner product
$
\lambda(x,t) = \langle x(t), w(t)\rangle  = w_0(t)+\sum_{i=1}^d x_i(t)w_i(t). 
$
This is still an extremely rich class of functions as $x_i(t)$ can be different over time and $w_i(t)$ is allowed to be any univariate nonnegative functions
over continuous time.
Furthermore, based on our intuition, $\lambda(x,t)$ is not a smoothly changing
function, but can jump suddenly in response to certain events.
It may not have a small or even bounded Lipschitz constant.
We therefore constrain complexity of the function class via total
variations (TV). Then we can learn the model by solving the variational penalized maximum likelihood problem below:
\begin{equation}
\begin{aligned}\label{eq:variational}
\min_{(w_0,w_1,...,w_d) \in \cF^{d}} &\sum_{i=1}^n\ell(\{x_i, z_i, \psi_i\}; {w}) + \gamma\sum_{j=0}^d \TV(w_j) \\
\text{s.t. } &w_j(t+\delta) -w_j(t)\geq 0 \text{ for any } j\in [d],t\in \R, \delta\in \R_+ 
\end{aligned}
\end{equation}
where $z_i$ is the indicator of censoring type for observation $x_i$, i.e.
interval-censored or right-censored; $\psi_i := \{t_i, T_i, T\}$ 
is the associated censoring time; $w_j(t)$ is the evaluation of
function $w_j$ at time $t$.
Note that the monotone constraint is optional and can be removed
to form a ``non-monotone'' model.
There, only issue is that \eqref{eq:variational} is an infinite dimensional function optimization problem and could be very hard to solve.

\subsection{Variational characterization}

The following theorem provides a finite set of simple basis functions
that can always represent at least one of the solution to
\eqref{eq:variational}.

\begin{theorem}[Representer Theorem]\label{thm:repre}
  Assume no observations are uncensored, feature $x_i(t)$ for
  each user is piecewise constant over time with finite number
  of change points. Let $s_{\tau}(t) = 1(t\geq \tau)$ be the
  step function at $\tau$. There exists an optimal solution
  $(w_1,...,w_d)$ of the above problem such that for each
  $j=1,..,d$, $$w_j(t) = \sum_{\tau\in \cT}^n s_{\tau}(t)c_{\tau}^{(j)}$$ 
	\vspace{-1em}
	
	for some set $\cT$ that collects all censoring boundaries and places where feature $x_j(t)$ changes, and coefficient vector $c^{(j)}\in \R^{|\cT|}$.
\end{theorem}

The direct consequence of Theorem~\ref{thm:repre} is that we can now represent piecewise constant functions by vectors in $\R^{|\cT|}$ and solve \eqref{eq:variational} by solving a tractable finite dimensional fused lasso problem (with an optional isotonic constraints) of the form:
\begin{equation}\label{eq:finite_optim}
\begin{aligned}
\min_{w_0,w_1,...,w_d \in \R^{|\cT|}} &\sum_{i=1}^n\ell(\{x_i, z_i, \psi_i\}; {w}) + \gamma\sum_{j=0}^d \|D w_j\|_1 \\
\text{s.t. } &w_j(\ell+1) -w_j(\ell)\geq 0  \text{, for any } j=1,...,d, \,\,\, \ell = 1,...,|\cT|-1. 
\end{aligned}
\end{equation}
where we abuse the notation to denote $w_j$ as evaluations of function $w_j$ at sorted time points in $\cT$; and $D\in \R^{(|\cT|-1)\times |\cT|}$ is the discrete difference operator.

\section{Experiments}
We evaluate the out-of-sample predictive power measured by
log-likelihood and conduct case studies of learned latent
hazard rate on real data via
domain expert's knowledge.

\subsection{Real-World Data}

The data used for evaluation was sourced from the work of Soska et al. ~\cite{Soska2014} and was compromised as a collection of interval censored sites from backlists and right censored sites randomly sampled from .com domains~\footnote{A .com zone file is the list of all registered .com domains at the time.}.

All the sites were drawn from The Wayback Machine\footnote{The Wayback Machine is a service that archives parts of the web.} when archives were available at appropriate dates. A total of \emph{49,347} blacklists were collected between 2011 to 2013,
include a blacklist of predominately phishing\footnote{A phishing website is a website that impersonates another site such as a bank, typically to trick users and steal credentials.}, and search redireciton attacks~\cite{Leontiadis2014}. The .com
zone files during the same period are randomly sampled, served as right censored sites with a total of \emph{336,671}.

We automatically extracted raw tags and attributes from webpages, that served as features (a total of {\em 159,000} features). These tags and attributes could be like <br>, or <meta> WordPress 2.9.2</meta>. Our corpus of features corresponds to a very large set of distinct and diverse software packages or content management systems.

\subsection{Real-World Numeric Results}

\begin{figure}[tbh]
	\centering
	\includegraphics[width=0.45\textwidth,height=0.45\textwidth]{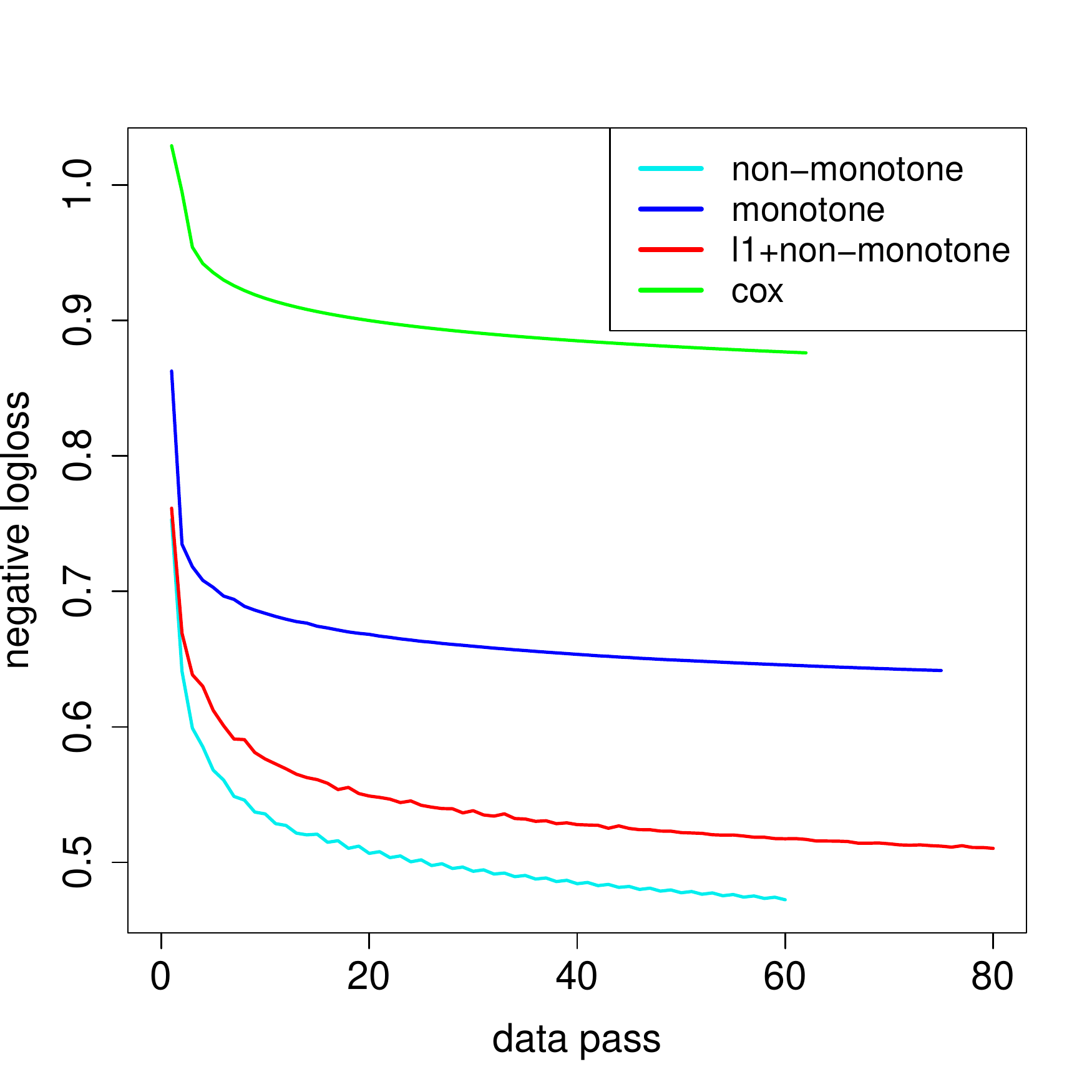}
	\includegraphics[width=0.45\textwidth,height=0.45\textwidth]{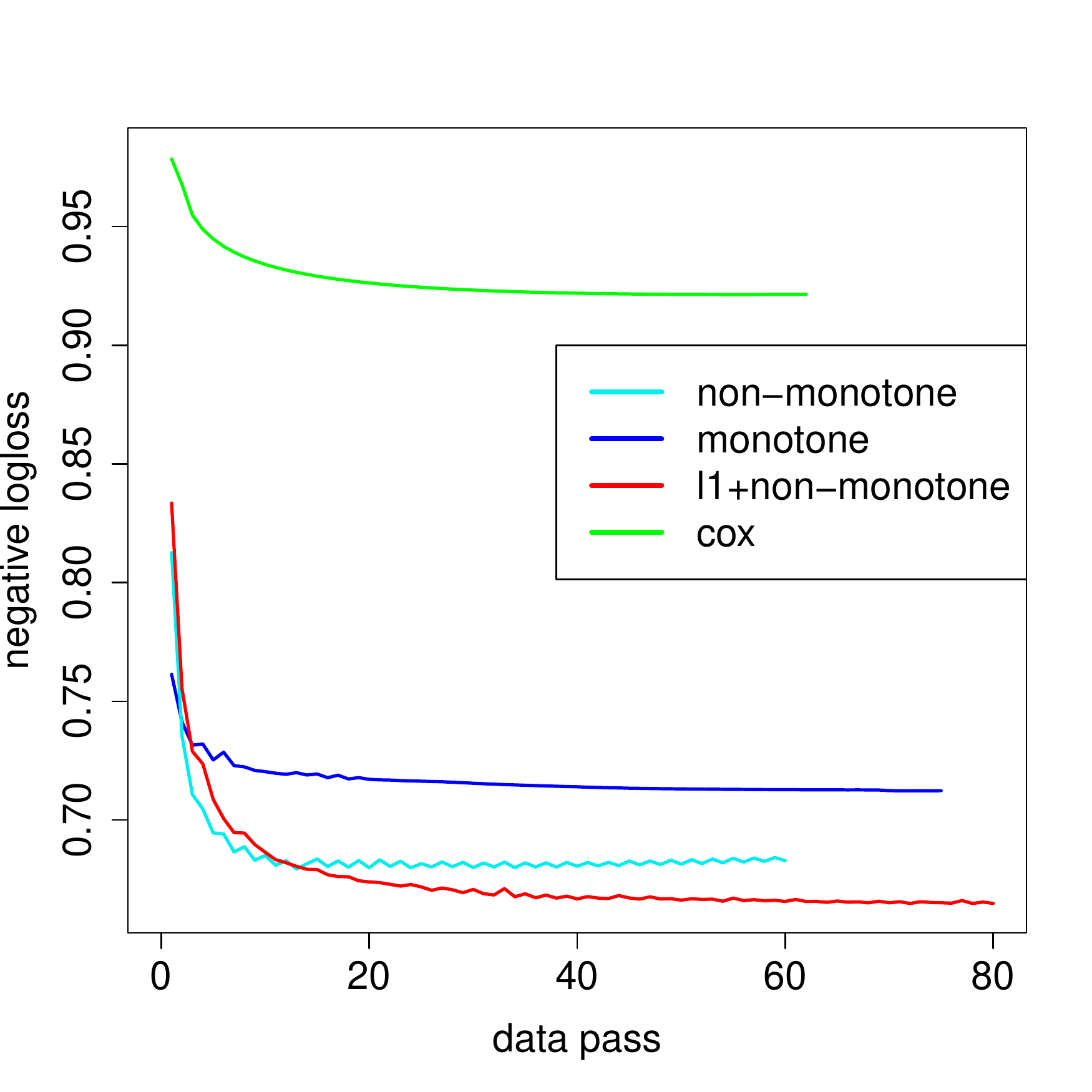}
	\caption{Convergence on training data ({\bf left}) and test data ({\bf right}) respectively.}\label{fig:real-convergence}
\end{figure}
The baseline method is the classic Cox Proportional model
(Cox)~\cite{Cox1972} which have been extensively used for
hazard regression and survival analysis. It's parametrized based on the 
features with constant coeffcients over time.
We use ``$\ell_1$'' to denote
the model penalized with $\|Dw^\top\|_1$.

An experimental comparison between our models and Cox on the aforementioned dataset are reported in Figure~\ref{fig:real-convergence} ({\bf left} and {\bf right}). Apparently the 
Cox model underfit the data quite a bit. Our ``monotone'' model allows only monotone hazard rate underfit the data a little but still significantly outperforms
Cox. It is well expected that ``non-monotone'' model without any constraint overfit the data severely. ``$\ell_1$+non-monotone''
model which is well regularized performs the best. This result clearly
shows that the latent hazard rate recovered by our model is much better than
the Cox's model.
To achieve such accuracy, our model uses only around 3 times parameter size compared with Cox model.

\begin{figure}[tbh]
	\centering
	\includegraphics[width=0.45\textwidth, height=0.45\textwidth]{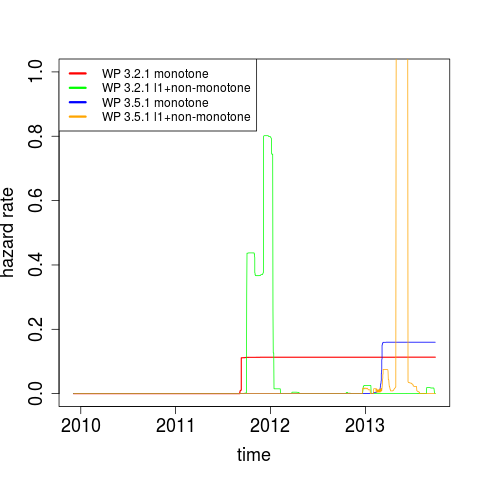}
	\includegraphics[width=0.45\textwidth, height=0.45\textwidth]{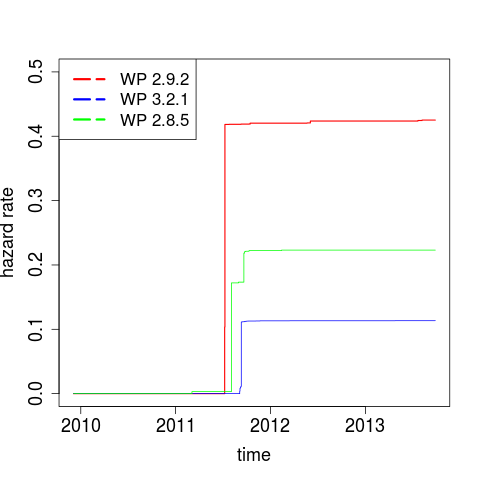}
	\caption{Estimated hazard rates (both ``monotone'' and ``non-monotone'') correspond to instances of Wordpress 3.2.1 and Wordpress 3.5.1 ({\bf left}).; Estimated hazard rates (``monotone'') correspond to different versions of Wordpress that were attacked in the summer of 2011 ({\bf right}).}\label{fig:real-case}
\end{figure}
\subsubsection{Real-World Case Study}\label{sec:case}



In this section, we manually inspect the model's ability to automatically discover
known security events. Figure~\ref{fig:real-case} ({\bf left}) demonstrates some of the differences between the monotone and non-monotone models by following the hazard assigned to features that correspond to Wordpress 3.5.1. In early 2013, our dataset recorded a few malicious instances of Wordpress 3.5.1 sites (among some benign ones). These initial samples appeared to be part of a small scale test or proof of concept by the adversary to demonstrate their ability to exploit the platform. Both models detect these security events and respond by assigning a non-zero hazard.

Following the small scale test was a lack of activity for a few weeks, during which the non-monotone model relaxes its hazard rate back down to zero, just before an attack campaign on a much larger scale is launched. This example illustrates
once a vulnerability for a software package is known, that package is always at risk, even if it is not actively being exploited. On the other hand, the non-monotone model captures the notion that adversaries tend to work in batches or attack campaigns. Previous work~\cite{Soska2014} has shown that it is economically efficient for adversaries to compromise similar sites in large batches, and after a few attack campaigns, most vulnerable websites tend to be ignored. This phenomena is shown in Figure~\ref{fig:real-case} ({\bf left}) where Wordpress 3.2.1 was attacked in late 2011 and then subsequently ignored with the exception of a few small attacks that were likely orthogonal to the underlying software and any observable content features. 

It can be observed from Figure~\ref{fig:real-case}~({\bf right}) that a number of distinct  Wordpress distributions experienced a change-point in the summer of 2011. This phenomina was present in several of the most popular versions of Wordpress in the dataset including versions 2.8.5, 2.9.2 and 3.2.1. This type of correlation between the hazard of features corresponding to different versions of a software package is expected. This correlation often occurs when adversaries exploit vulnerabilities which are present in multiple versions of a package, or plugins and third party add-ons that share compatibility across the different packages.

\section{Conclusion}
In this paper, we propose a novel survival analysis-based approach to model the latent process of websites getting hacked over time. The proposed model attempts to solve a variational total variation penalized optimization problem, and we show that the optimal function can be linearly represented by a set of step functions with the jump points known to as ahead of time. The results suggest that by correctly recovering the latent hazard rate, our model significantly outperforms the classic Cox model. Compared with known time-dependent hazard regression models, our models work on
several orders of larger feature space. Most importantly, identifying the changes of each feature's susceptibility over time can help people understand the latent hacking campaigns and leverage these insights to take appropriate actions.

In future, further works can be made to study the relations among potential
correlated features by investigating the structures (e.g. low rank) of the coefficient matrix $w \in \RR^{d\times \cT}$, or via deeper transformation. On the other hand,
the same model (variants) can be used in many other settings to study consumer
spending behaviors, marriage, animal habits and so on.

\bibliographystyle{icml2016}

\end{document}